\documentclass[3p,times]{elsarticle}

\usepackage{ecrc}

\usepackage{amsmath}

\usepackage{graphicx}

\usepackage{subcaption}

\usepackage{lmodern}

\usepackage{multirow, tabularx}

\usepackage{adjustbox}

\usepackage{enumitem}

\volume{00}

\firstpage{1}

\journalname{Journal Name}

\runauth{}

\jid{procs}

\jnltitlelogo{Journal Logo}

\usepackage{hyperref}

\usepackage{amssymb}

\usepackage[figuresright]{rotating}

\usepackage{makecell}

\begin{document}

\begin{frontmatter}

\dochead{}

\title{An Adaptive Image Encryption Scheme Guided by Fuzzy Models}

 \author{Mahdi Shariatzadeh}
 \author{Mohammad Javad Rostami\corref{cor1}} 
 \ead{Tel:+989133417396, Fax:+3432477400, Email address: mjrostamy@yahoo.com, rostami@uk.ac.ir}
 \cortext[cor1]{Corresponding author}
 \author{Mahdi Eftekhari}
 \address{Department of Computer Engineering, Shahid Bahonar University of Kerman, Kerman, Iran}

\begin{abstract}
A new image encryption scheme using the advanced encryption standard (AES), a chaotic map, a genetic operator, and a fuzzy inference system is proposed in this paper. In this work, plain images were used as input, and the required security level was achieved. Security criteria were computed after running a proposed encryption process. Then an adaptive fuzzy system decided whether to repeat the encryption process, terminate it, or run the next stage based on the achieved results and user demand. The SHA-512 hash function was employed to increase key sensitivity. Security analysis was conducted to evaluate the security of the proposed scheme, which showed it had high security and all the criteria necessary for a good and eﬀicient encryption algorithm were met. Simulation results and the comparison of similar works showed the proposed encryptor had a pseudo-noise output and was strongly dependent upon the changing key and plain image.
\end{abstract}


\end{frontmatter}

\section{Introduction}
\label{}

With the expansion of data analysis in recent years, data security has become particularly important. Data security is provided in different ways, according to the type of data, whether it is text, audio, image, video,  etc. Steganography, cryptography, and watermarking are some of the most common ways to secure multimedia data
\cite{roy2019study}.
In the field of cryptography, data is transformed into meaningless and incomprehensible formats in order to make the information unusable for everyone except the people who hold the cryptographic key
\cite{xu2021applications}.
In steganography, considering that the meaninglessness of information attracts the attention of others to the information and its users, the information is hidden inside another type of information called the cover
\cite{abdulla2019improving}.
Watermarking is done for different purposes such as copyright protection, data integrity, publishing path tracking, etc.
\cite{hurrah2019dual}.
The image, as one of the most commonly used data types, has always been one of the most attractive fields of research for researchers in various fields of science. The use of images in various fields such as art, the military, advertising, commerce, etc. has added to this attractiveness. Therefore, the importance of image security is undeniable. As mentioned, encryption is one of the ways to secure data, including image data.

Image encryption with the AES method is not effective when there is a strong correlation between pixels. This problem becomes more serious when the image has redundancy
\cite{wu2018image,zhang2018plaintext}.
Symmetric encryption algorithms such as AES, DES, etc. are performed on small data blocks, generally128 bits long. The encryptor output, which is replaced and sent, is equivalent to the input data block. The number of encryptor input and output data blocks is similar. For example, if block m is repeated in plaintext a hundred times, the block it is replaced with will be viewed a hundred times in cipher text. Therefore for image blocks that are identical, the equivalent replaced block will be repeated in different sections of the image. Consequently uniform sections of the image such as the image background will remain unchanged after the encryption, hence the whole image is seen at a single glance
\cite{zhang2018test}. 
Although AES is presently a good method for text encryption, it is not suitable for image encryption due to the following reasons
\cite{zhang2017fast}:
\begin{enumerate}[label=(\alph*)]
	\item
	The high volume of images tends to lower the speed of the algorithm.
	\item
	Pixel correlation between adjacent pixels is reflected in the encrypted image.
	\item
	Plain image redundancy is reflected in cipher image.
\end{enumerate}
Therefore, it is crucial to make modifications to standard encryption algorithms to improve the image features for use in image encryption algorithms.

Chaotic maps are finite and deterministic while showing complex and dynamic behavior. Accordingly, chaotic encryption and its eﬀiciency in image encryption is better than common symmetric methods because of the image size, correlation between adjacent pixels, and redundancy in images
\cite{radwan2016symmetric};
furthermore there is a strong relationship between the dynamic specification of chaotic maps and the required properties for the encryption system, which allows for the use of chaotic maps in image encryption
\cite{tang2019delay}.

The rest of this paper is organized as follows. The next section discusses related works. Section \ref{sec:prp} describes the proposed algorithm and related preliminaries. Section \ref{sec:result} includes results and the security analysis of the proposed algorithm. Finally, section \ref{sec:conc} concludes the paper.

\section{Related Works}

\subsection{Image encryption methods using chaotic maps}
The use of chaotic maps has many applications in image encryption. Arab et al.
\cite{arab2019image}
presented a new algorithm using the Arnold chaos sequence, integrating it with a modified version of AES. Rostami et al. 
\cite{rostami2017novel}
presented an algorithm with acceptable results using the logistic chaotic map, the XOR operator, and chaotic windows. They stated that their proposed algorithm could be employed in noisy communication networks and parallel processing.

Wang et al. 
\cite{wang2015fast}
were able to encrypt images with acceptable security using a logistic map and switching rows and columns.
Liu et al.
\cite{liu2016fast}
introduced a new chaotic map called 2D-SIMM. The map is based on the sine map and iterative chaotic map with infinite collapse (ICMIC). They evaluated the eﬀiciency of their chaotic map by several assessment methods and succeeded in introducing 2D-SIMM as a suitable chaotic map for image encryption. In addition, based on 2D-SIMM, an algorithm for image encryption was introduced with high security, low time complexity, and resistance to brute-force, statistical, known-plaintext, differential, and chosen-plaintext attacks. In fact, the algorithm introduced in 
\cite{liu2016fast}
is obtained by improving the key space in 
\cite{wang2015fast}.

Pak and Huang
\cite{pak2017new}
proposed a method for encrypting color images using the output sequence differences of two one-dimensional chaos maps. One of their innovations was the introduction of a new chaos map and a three-step algorithm for image encryption. The three main steps in the algorithm proposed by 
\cite{pak2017new}
are by a new permutation related to the key generator module, a diffusion that is also related to the key generator module, and a linear transformation whose output creates the final encrypted image. Li et al.
\cite{li2017hyper}
proposed another algorithm for image encryption, using a similar architecture and a 5D chaotic map. The algorithm proposed in
\cite{li2017hyper}
performs permutation operations at both pixel and bit levels. They tried to solve the common problems in image encryption algorithms using a high-dimensional chaotic map.

Li et al.
\cite{li2017image}
proposed an encryption scheme for images based on a chaotic tent map. Their scheme is done in two steps. In the first step, a key for cryptography is generated using a chaotic tent map. In the second step, the encryption operation is performed using the key created in the first step. Investigations have shown that their scheme does not provide adequate security.
Hermassi et al.
\cite{hermassi2013improvement} 
introduced an enhancement on the hyper-chaotic image encryption scheme. Ye et al.
\cite{ye2012efficient}
proposed an algorithm for image encryption based on the generalized Arnold map.
Norouzi et al. 
\cite{norouzi2014simple},
using hyper-chaotic sequences and plaintext, introduced a new algorithm for encrypting images, which worked well.
Song et al. 
\cite{song2013image}
introduced a new spatiotemporal chaotic system by defining a nonlinear map in coupled map lattices (CML) and a chaotic map. Shariatzadeh et al. 
\cite{shariatzadeh2021proposing}
proposed an AES-inspired algorithm for image encryption using logistic chaotic map and 256 values in the finite field $2^{8}$.

\subsection{Image encryption methods using DNA}
There are methods and tools other than using chaotic maps to encrypt images. Another way is to use deoxyribonucleic acid (DNA) technology to encrypt images. DNA molecules contain genetic information. In this category of image encryption methods, an attempt is made to use DNA molecules' computational power by simulating them
\cite{kaur2020comprehensive}.
Li et al.
\cite{li2016improvement}
developed a new algorithm for image encryption using DNA, XOR, and real and complex turbulent systems. They first created three sequences from color images using coding on R, G, and B channels. They then created equal blocks using the Hamming distance between the two generated keys. Finally, using the XOR operator, they combined the generated blocks with DNA codes. The results obtained in 
\cite{li2016improvement}
show acceptable resistance to statistical and differential attacks.
Wu et al.
\cite{wu2018image}
also introduced a new method for encrypting images by introducing a new chaotic map called the two-dimensional Henon-Sine map (2D-HSM) using DNA encoding. Another method that has used the DNA model is the method suggested by Enayatifar et al. 
\cite{enayatifar2017image}.
They first turned two-dimensional images into one-dimensional ones. To increase the speed, they performed the permutation and diffusion steps together. To perform these two steps, they used chaos mapping and DNA coding.

Wu et al.
\cite{wu2015new}
presented a four-step scheme for color images based on three chaotic one-dimensional
systems and DNA sequencing operations. In the first step, the cryptographic keys are prepared. This step is done using three one-dimensional chaotic maps and the plain image. The cryptographic keys are then converted to DNA matrices using DNA coding rules. In the second step, using the XOR operation, the DNA matrices are changed. In the third step, the DNA matrices are divided into equal blocks and these blocks are randomly scrambled. In the fourth step, using DNA decoding rules, DNA matrices are converted into an encrypted image. The results presented in
\cite{wu2015new}
indicate that this algorithm has a high resistance to statistical and differential attacks. The algorithm proposed by
\cite{wu2015new}
also deals with the noise resistance of the cryptographic system. The results show that this algorithm has a higher resistance to salt and pepper noise than to Gaussian noise. DNA encoding and computation modulo 4 have been used to reduce the complexities of the image encryption problem 
\cite{rostami2016chaos}.

\subsection{Image encryption methods using cellular automata}
Another approach to image encryption is the use of cellular automata. Cellular automata are made up of cells that are placed on a network with different structures. These structures evolve in a variety of ways and simulate complex structures. The vast body of rules in cellular automation is also developed in a variety of ways. This evolution is created through simple logical calculations with random behaviors. Reversible cellular automation is widely used for block encryption. The main advantages of cellular automation in encryption are the large space of rules and parallel work
\cite{kaur2020comprehensive}.

Ping et al.
\cite{ping2014image}
used a balanced and non-aﬀine type of cellular automata with complex behavior as blocks for their image encryption scheme. In their scheme, local interaction between cells performs diffusion operations, while confusion steps are achieved by nonlinear rules. With the method used in
\cite{ping2014image},
the confusion and diffusion stages are well integrated. The cryptographic scheme of Niyat et al.
\cite{niyat2017color}
is based on cellular automata. Using a uniform cellular automata framework, they tried to overcome the problems that this set of cells have. They changed the position of the pixels using chaotic maps and created an image as a key using cellular automata.
\subsection{Image encryption methods using Meta-heuristic algorithms}
Meta-heuristic algorithms are also used in the field of image encryption, and they are based on population behavior. Meta-heuristic techniques are suitable for solving NP-hard and optimization problems, and these techniques can be used to optimize static parameters in image encryption algorithms
\cite{kaur2020comprehensive,enayatifar2014chaos}.
Using dynamic harmony search (DHS) and tent chaos mapping, Talarposhti and Jamei
\cite{talarposhti2016secure}
proposed an algorithm for encrypting grayscale images. The encryption process in their proposed algorithm is done in two steps. In the first stage, which is the diffusion stage, the fitness function is maximized using DHS. They used entropy as a fitness function. In the second stage, permutation is applied horizontally and vertically. One of the advantages of their proposed algorithm is low time complexity.
\subsection{Image encryption methods using fuzzy logic}
The fuzzy-based approach is another technique for image encryption 
\cite{kaur2020comprehensive}.
Fuzzy logic is used in fuzzy systems. Fuzzy logic is a form of a multi-valued region in which the logical value of variables can be any real number between 0 and 1. This logic is used to apply the concept of partial correctness. In this logic, the amount of correctness can be any value between completely true and completely false
\cite{de2021interpretable}.
Seyedzadeh et al.
\cite{seyedzadeh2014rgb}
have presented a color image encryption based on the Choquet fuzzy integral (CFI), which generates pseudo-random keystreams. The output of the CFI is used to randomly shift the bits of three gray-level components; then the three components of RGB color pixels and the generated keystream are coupled to encrypt the permuted components.

Qaid and Sanjay
\cite{qaid2013encrypting}
have supported user-desired security levels and their corresponding processing levels by generating random keys for the encryption/decryption process; this is achieved by using fuzzy logic and fuzzy sets.
Hashemi
\cite{hashemi2013design}
introduces an image encryption algorithm based on DNA addition combining and a coupled two-dimensional piecewise nonlinear chaotic map. The generated sequence of the Sugeno fuzzy integral and the DNA sequence addition operation is used to add encoded color component blocks.

\subsection{Other related works}
For real time applications, Bahrami and Naderi 
\cite{bahrami2012image}
have proposed a stream encryption algorithm that has a good performance when faced with high speed and the error probability of data transmission. The results have showed that the performance speed of the proposed algorithm is better compared to A5/1 and w7 stream ciphers.
Kumar et al. 
\cite{kumar2011comments}
have performed a cryptanalysis on one of the image encryption algorithms and showed that it is vulnerable to some types of attack.

In this paper, in order to use an AES algorithm in image encryption, a chaotic tent map, whose properties will be discussed in detail later in this paper, is utilized as a simple chaos map. A random block is generated with a tent chaotic map the size of AES-128 encryption blocks, which was XORed with the first plain image block before running the AES. The genetic cross over operator is used to increase the resistance of the proposed algorithm against differential attacks, and finally, in case of process stop or continuation, a proposed adaptive fuzzy system decides the type of input image and the results of the previous sections according to user requirements.

The contributions of this paper include:

\begin{itemize}
	\item
	Using fuzzy models to provide security for the encrypted image according to user requirements and input images.
	\item
	Using a crossover genetic operation for changing the gray level of each pixel of the image.
	\item
	Using a chaotic logistic map combined with an AES algorithm to overcome the drawbacks of a block- based encryptor for the image encryption problem.
	\item
	Using the SHA-512 hash function to extract the needed key for the AES-128 from a 1029 bit input master key.
	\item
	Using an eﬀicient method for pixel shuffling using the chaotic map.
	\item
	Increasing the key sensitivity through obtaining dependent initial conditions upon the SHA-512 output key for the tent map.
	\item
	The diffusion property is increased through using the sum of the last encrypted block as a factor in the necessary initial value for the tent map, which leads to increased resistance against differential attacks. 
\end{itemize}

\section{Proposed Algorithm}
\label{sec:prp}
One of the problems of most AES-based encryption algorithms is their block-based structure. In this structure, due to the separate encrypting of each block, little change (one block or a few pixels) in the plain image will lead to very little change in the encrypted image indicating inappropriate diffusion. In addition, in most of these structures, permutation is conducted at the block level only while in the proposed algorithm, permutation is applied to the entire image. Furthermore, since the AES-related algorithms have block-based structures, if the plain image is a uniform image, encrypted blocks will be quite similar, leading one to recognize the uniform image. 

In the proposed algorithm, in addition to the elimination of the mentioned drawbacks in a combinational module, an adaptive fuzzy system decides on continuing the encryption process by using a genetic crossover operation, repeating  the combinational module, or terminating the process due to the achievement of results based on user demand. The proposed system has been presented in two phases after the description of preliminaries.

\subsection{Preliminaries}
This section gives a brief review on  the preliminary materials for the proposed algorithm.

\subsubsection{Tent chaotic map}
Chaotic mapping in image encryption is very common and widely used
\cite{xiang2020improved}.
Chaos mappings are nonlinear dynamic systems that are very sensitive to their initial conditions. So far, several chaotic maps have been proposed. Each of them exhibits different behaviors and properties. One of the simplest of these mappings is tent chaos mapping. The tent Chaos map was introduced by Yoshida in 1983
\cite{naskar2020robust}.
The tent map has a parameter called
$\mu$
and is one-dimensional. Tent map is defined by
\begin{equation}
	x_{t+1}=
	\begin{cases}
		\mu x_{t}  & x_{t}<\dfrac{1}{2}\\
		\\
		\mu (1-x_{t})  & x_{t} \ge \dfrac{1}{2}.
	\end{cases}	
\end{equation}

\begin{figure*}[t]
	\centering
	\begin{subfigure}[b]{0.5\linewidth}
		\includegraphics[width=\linewidth]{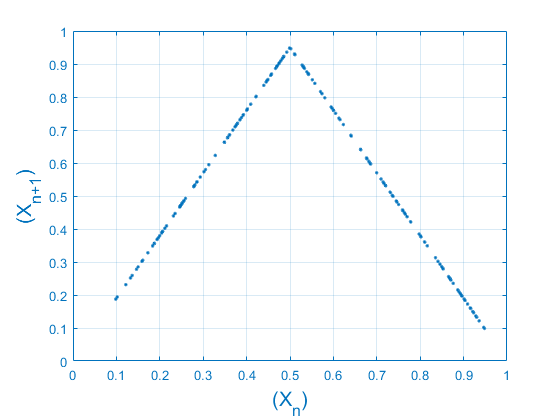}
		\caption{The Tent chaos map path}
		\label{fig:tent_a}
	\end{subfigure}
	\begin{subfigure}[b]{0.43\linewidth}
		\includegraphics[width=\linewidth]{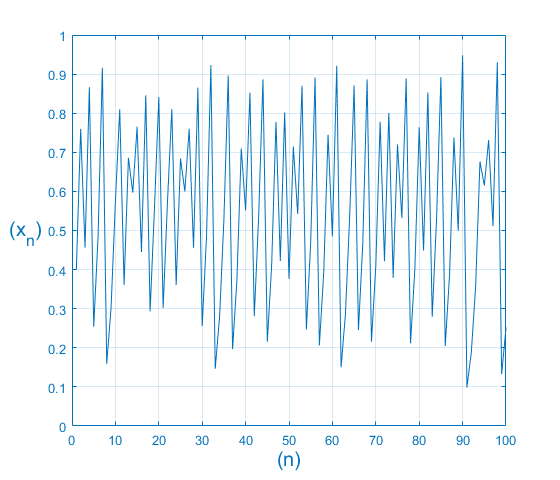}
		\caption{Tent map with $\mu$=1.9}
		\label{fig:tent_b}
	\end{subfigure}
	\begin{subfigure}[b]{0.32\linewidth}
		\includegraphics[width=\linewidth]{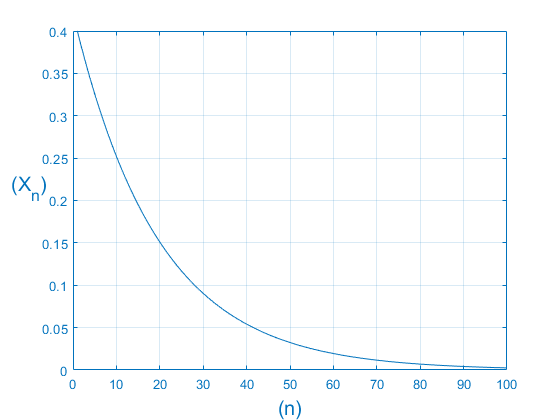}
		\caption{Tent map with $\mu$=0.95}
		\label{fig:tent_c}
	\end{subfigure}
	\begin{subfigure}[b]{0.32\linewidth}
		\includegraphics[width=\linewidth]{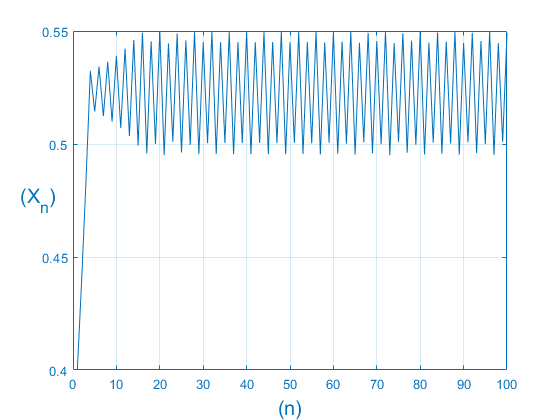}
		\caption{Tent map with $\mu$=1.1}
		\label{fig:tent_d}
	\end{subfigure}
	\begin{subfigure}[b]{0.32\linewidth}
		\includegraphics[width=\linewidth]{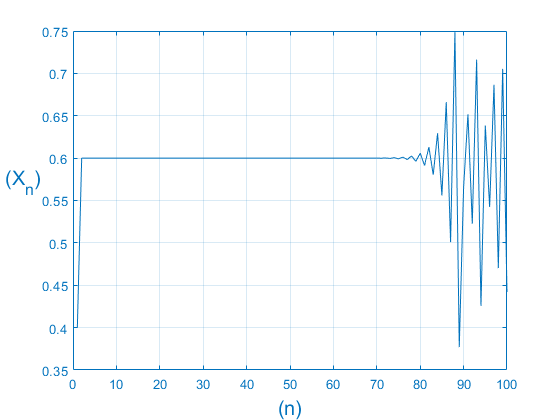}
		\caption{Tent map with $\mu$=1.5}
		\label{fig:tent_e}
	\end{subfigure}
	\caption{Tent map}
	\label{fig:tent}
\end{figure*}
Figure \ref{fig:tent_a} shows the tent chaos map path.
Figures \ref{fig:tent_b}, \ref{fig:tent_c}, \ref{fig:tent_d}, \ref{fig:tent_e} show the tent chaos map behavior for different $\mu$ values. In these graphs, the initial value of $x_{t}$ is assumed to be 0.4.

\subsubsection{Pixel Shuffling}
\label{sec:pixelshuffling}
Pixel shuffling increases the security level of the image encryption algorithm 
\cite{raza2019novel}. 
This operation is implemented in various methods along with chaotic maps. In the proposed algorithm, pixel shuffling was used with simple types of such methods. This procedure includes a gray-scale image with a $M\times N$ size ($Im_{M\times N}$). The image is converted and then the vector is achieved 
($P_{1 \times (MN)}=\{P_{1}, P_{2} ... P_{MN}\}$); Next, the tent chaotic map is iterated with the initial value $x_{0p}$ and the parameter $\mu_{p}$ to generate a random sequence of length $M\times N$. By sorting them in ascending order and getting their indices, $I_{1 \times(MN)}=\{i_{1}, i_{2} ... i_{MN}\}$ is generated. Pixel shuffling is normally used to reorder $P_{1 \times (MN)}$. For example, the $i_{1st}$ and $i_{2nd}$ elements of $P_{1 \times (MN)}$, reside in the first and second position of the reordered vector. The reverse shuffling operation runs with sequence $I$. For example, the first and the second elements of $P_{p}$ reside in the $i_{1st}$ and $i_{2nd}$ place of $P$.

\subsubsection{Genetic Crossover Operator}
\label{sec:crossover}
The cross over genetic operation is used for changing the gray level of each pixel of the image and the chaos map by 
\subparagraph{Step 1}
Initializing the tent chaotic map: To increase the security of the proposed method, this value is specified from the 512-bit hash code of the key.
 
\subparagraph{Step 2}
Selecting two pixels of the image for cross over operation: The first pixel: At the beginning of the process, the image is converted to 1D matrix that has one row and 65536 columns. The first pixel is selected (in the next step, the second pixel is selected, and then the sequential pixels to the last pixel are selected).The second pixel: Chaos map is utilized to select the second pixel.
 
\subparagraph{Step 3} 
Using the selected pixels for the crossover operation: The two specified pixels in the previous step are selected for the crossover operation after converting the gray level to base 2. The used cross over operation in this stage is a single-point crossover.

\subparagraph{Step 4}
Steps 2 and 3 are repeated for all pixels of the image.

\subparagraph{}
The decryption procedure is carried out in the reverse order of the encryption procedure.

\subsubsection{SHA-512}
\label{sec:shasection}
\begin{figure*}[t]
	\centering
	\captionsetup{justification=centering}
	\includegraphics[scale=0.8]{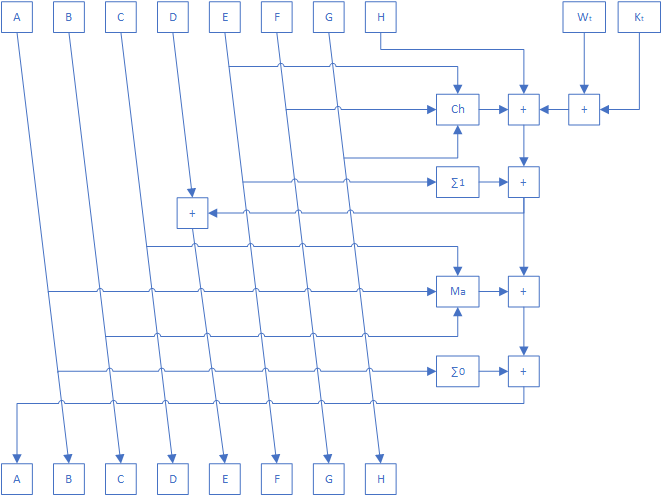}
	\caption{One iteration of the SHA-512 hash function. The components perform the following operations:
		\\
		$Ch(E, F, G) = (E \wedge F) \oplus (\neg E \wedge G), Ma(A, B, C) = (A \wedge B) \oplus (A \wedge C) \oplus (B \wedge C),
		\newline
		\sum0(A) = (A>>>2) \oplus (E>>>13) \oplus (A>>>22), \sum 1(E) = (E>>>6) \oplus (E>>>11) \oplus (E>>>25)$,
		\newline
		 The + is addition modulo $2^{64}$}
	\label{fig:sha}
\end{figure*}
The SHA-512 is a hash algorithm for cryptography that can present the message digest in 512-bit format
\cite{saravanan2021hap}. 
The National Institute of Standards and Technology (NIST) released SHA-2 as the information processing standard in 2001. This was done due to the detection of a security problem in the SHA-1 and the second version of SHA tried to fix the problems of its previous version. SHA-2 was designed by the US National Security Agency (NSA) and actually included SHA-224, SHA-256, SHA-384 and SHA-512. The end number of these algorithms indicates the output length of these algorithms.

The proposed algorithm for image encryption in this paper uses SHA-512. SHA-512 processes 64-bit words in 80 rounds. The SHA-512 input size consists of eight 64-bit words. Figure 2 shows the integration performance in each round of the SHA-512 algorithm. As can be seen in this figure, the hash operation by SHA-512 is performed by the AND, XOR, Rotation, ADD, and OR operators.

\subsubsection{AES}
\label{sec:aessection}
\begin{figure*}[t]
	\centering
	\includegraphics[scale=0.9]{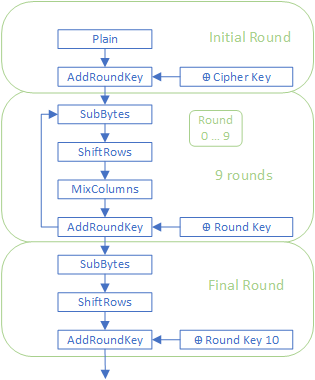}
	\caption{Flowchart of AES-128}
	\label{fig:aes}
\end{figure*}

\begin{table}[b]
	\centering
	\caption{Data block length, key length and number of iterations in the AES algorithm}
	\begin{tabular}{c|c|c|c|}
		\cline{2-4}
		& Key length & Data block length & Number of iterations \\ \hline
		\multicolumn{1}{|c|}{AES-128} & 4          & 4                 & 10                    \\ \hline
		\multicolumn{1}{|c|}{AES-192} & 6          & 4                 & 12                    \\ \hline
		\multicolumn{1}{|c|}{AES-256} & 8          & 4                 & 14                    \\ \hline
	\end{tabular}
	\label{tab:aes}
\end{table}
NIST published the AES in the FIPS 197 document in 2001
\cite{kumar2018fpga}. 
AES is a symmetric cryptographic algorithm. This means that the encryption and decryption operations are performed with the same key. In AES, the length of a data block is 128 bits (equivalent to four 32-bit words). The encryption key length can be 128 bits, 192 bits, or 256 bits. Selecting any of these keys causes the number of iterations of the algorithm to run differently. Table \ref{tab:aes} shows the length of the data block, the length of the key, and the number of iterations in the AES algorithm.

The description of the AES algorithm is very simple, but the mathematical foundations of this algorithm are very detailed. The AES performs encryption operations using four functions: AddRoundKey, SubBytes, ShiftRows, and MixColumn. The AES also expands the key using a complex and irreversible algorithm. The flow chart of the AES 128 algorithm is shown in Figure \ref{fig:aes}.

In the SubByte function, each of the input bytes is replaced with new values based on a fixed S-Box. This S-Box is a matrix that has 16 rows and 16 columns. In the RotateRows function, three input rows (which is a $4 \times 4$ matrix) rotate to the left. In the AddRoundKey function, each input item is XORed byte by byte with the corresponding item from the key in that iteration. In the MixColumn function, each column of the input is integrated and tangled independently of the other columns in the finite field ($2^{8}$).

\subsubsection{Fuzzy Inference Systems}
\begin{figure*}[b]
	\centering
	\includegraphics[scale=0.8]{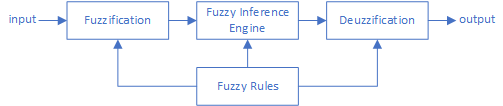}
	\caption{A scheme of a fuzzy inference system}
	\label{fig:fis}
\end{figure*}
Fuzzy logic based on fuzzy sets was first introduced by Professor Lotfi A. Zadeh in 1965
\cite{mendel2020human}.
This logic was developed and put into practice in 1970. In fuzzy sets, variables are linguistic. The value of linguistic variables is not a number, but words or sentences of a natural or artificial language. Unlike Aristotelian and binary logic, where all variables are considered as yes or no or zero or one, in fuzzy logic the variables change in the range between zero and one.
\cite{de2021interpretable}.
fuzzy inference system performs inference operations based on fuzzy sets and fuzzy rules. The fuzzy inference system consists of three components named fuzzification, fuzzy inference engine, and defuzzification with the architecture shown in Figure \ref{fig:fis}.

Numerical variables are converted to linguistic variables by fuzzification. In the inference engine, fuzzy rules are evaluated and inferred by inference algorithms. The output of this section is given to defuzzification to convert the generated values to a numeric value again. Defuzzification methods include the weighted average, centroid, center of sums, center of largest area, first of maxima, last of maxima, max membership, and mean-max membership methods 
\cite{na2012new}.

Fuzzy inference systems come in many forms. One of these systems is the Mamdani fuzzy inference system. This system has high expressive power and is implemented in two ways: multiple input multiple output (MIMO) and multiple input single output (MISO). This system uses fuzzy sets as a result of rules. The output of the rules in this system is nonlinear and fuzzy
\cite{salman2010comparison}.

\subsection{Encryption Process}
The proposed algorithm encrypts the images in two phases. In the first phase, using three operators XOR, pixel shuffling, and genetic crossover, the confusion and diffusion of the system are increased. Increasing diffusion means that the cryptographic system must prevent the transfer of the statistical features of the input image to the output of the system. In other words, the output pixels should not be correlated. Increasing confusion also means that there should be no relationship between the input image and the encrypted image.

In the second phase, using fuzzy inference systems, an attempt is made to increase the static diversity (S-Dive) and dynamic diversity (D-Dive) of the image. Static diversity refers to the statistical indicators of an image, which are measured by entropy in this paper. Dynamic diversity also refers to indicators related to avalanche effects and differential attacks, which are measured using the UACI and the NPCR. The avalanche effect in cryptographic systems means that a very small change in the input must lead to a very large and unpredictable changes in the output.

The description of the encryption phases is by

\paragraph{Phase 1}
This phase is done in 5 steps by

\subparagraph{Step 1}
Using SHA-512 the hash of the plain image is calculated. This hash code is used for encryption operations in the second phase as well as for decryption operations. After the encryption operation, this hash code is appended to the primary key.

\subparagraph{Step 2}
The SHA-512 hash of the 512 bit Key is computed to achieve a 512 bit sequence ($K_{512}$) that depends on the primary key, then generate the new initial conditions for the tent chaotic map and another key for the next phase by following the equations:

\begin{equation}
	\label{eq:key}
	\begin{array}{l@{}l}
		x_{0}=\dfrac{bi2de(K_{512}[1:128])}{2^{128}}, \\
		\\
		x_{1}=\dfrac{bi2de(K_{512}[129:256])}{2^{128}}, \\
		\\
		x_{2}=\dfrac{bi2de(K_{512}[257:384])}{2^{128}}, \\
		\\
		2nd\;phase\;key=K_{512}[385:512],
	\end{array}
\end{equation}
where '$bi2de$' converts a binary row vector to a non-negative decimal integer. Since the generated 512 bit hashed key is dependent on the primary key, if one bit is changed in the primary key, the hashed key and x0, x1, and x2 will be changed dramatically, which strengthens the diffusion property.

\subparagraph{Step 3}
Using the tent chaos map with the initial value of $x_{0}$ obtained from the primary key in the previous step, a sequence equal to the number of pixels in the plain image is created. The values obtained are in the range $[0, 1]$ and are doubled. Using the following equation, these values are mapped to integers and non-negative numbers in the range $[0, 255]$:

\begin{equation}
	t_{i}=mod(x_{i} \times 10^{14}, 256),
\end{equation}
where $x_{i}$ is the value created by the tent chaos map and $t_{i}$ is an integer obtained in the range $[0, 255]$. In this step, after creating the sequence $T$ where $T=\{0 \le t_{i} \le 255, t_{i} \in \mathbb{Z} \}$, each of the pixels of the plain image with its corresponding number in the sequence $T$ is XORed bit by bit in binary form.

\subparagraph{Step 4}
To increase diffusion and confusion in the cryptographic system, pixel shuffling is applied to the image as described in Section \ref{sec:pixelshuffling}. The initial value for the tent chaos map in this step is $x_{1}$.

\subparagraph{Step 5}
\begin{figure*}[b]
	\centering
	\includegraphics[scale=1]{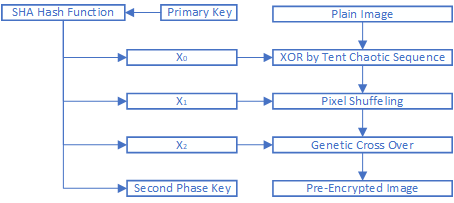}
	\caption{Architecture of encryption process in the first phase}
	\label{fig:first}
\end{figure*}
The genetic crossover operation is applied to the image as described in section \ref{sec:crossover}. As in the previous step, this is done to increase confusion and diffusion in the cryptographic system. The initial value for the tent chaos map in this step is $x_{2}$. The output of this step is an image called the pre-encrypted image.

Comparison of the plain image and pre-encrypted images has shown that steps 3 to 5 significantly eliminate the statistical properties of the plain image and increase confusion and diffusion in the encryption system. This operation causes the system to show high resistance to statistical attacks. The flowchart of the first phase of the cryptographic system is shown in Figure \ref{fig:first}.

\paragraph{Phase 2}
\begin{figure*}[t]
	\centering
	\includegraphics[scale=1]{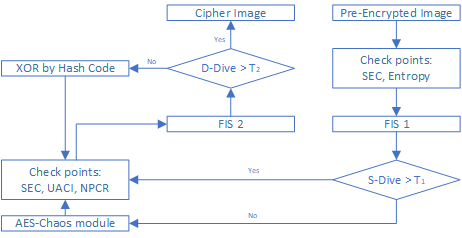}
	\caption{Architecture of encryption process in the second phase}
	\label{fig:second}
\end{figure*}
The encryption process on the pre-encrypted image is performed in the second phase with the aim of increasing the static and dynamic diversity under the supervision of two fuzzy inference systems named FIS1 and FIS2. In this phase, two modules called ”AES Chaos Module” and ”XOR by hash code”are used for encryption. These two modules use the AES and SHA-512 as described in sections \ref{sec:shasection} and \ref{sec:aessection}. 
This phase also uses two check points for encryption. At the first check point, the pre-encrypted image entropy value is calculated and sent to FIS1 along with the SEC value, which indicates the level of security requested by the user. The function of FIS1 is to calculate the value of S-Dive. If the value of S-Dive is less than the value of $T_{1}$, the image is sent to the AES-Chaos module. Studies have shown that the AES-Chaos module increases static diversity. If the value of S-Dive is greater than the value of $T_{1}$, the AES-Chaos module will not be applied to the image. If AES-Chaos Module is applied to the image, a bit with a value of one is added to the decryption key and otherwise a bit with a value of zero is added.

At the second check point, the UACI and NPCR values of the image are calculated and sent to FIS2 with the SEC. The function of FIS2 is to calculate D-Dive. If the value of D-Dive is less than the value of $T_{2}$, the XOR operation is performed on the image by hash code. After performing XOR by hash code, the value of D-Dive is calculated again. To calculate D-Dive, it is necessary to recalculate the UACI and NPCR values of the image. If the D-Dive value is again less than the $T_{2}$ value, the XOR is performed on the image by hash code operation again. This process is repeated until either D-Dive exceeds the value of $T_{2}$ or finally the operation XOR by hash code is applied to the image fifteen times. Studies have shown that applying XOR by hash code more than 15 times not only does not increase D-Dive, but in some cases, even reduces D-Dive. At the end of the encryption operation, four bits are added to the decryption key. The value of these four bits is between zero and 15 and indicates the number of XOR by hash code operations applied to the image.

\begin{figure*}[b]
	\centering
	\begin{subfigure}[b]{0.325\linewidth}
		\includegraphics[width=\linewidth]{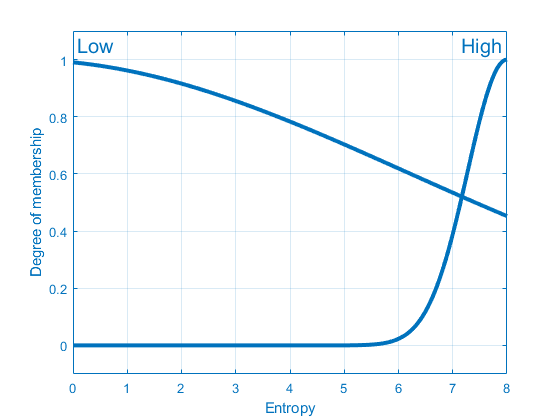}
		\caption{Membership functions for Entropy input}
		\label{fig:fis1_ent}
	\end{subfigure}
	\begin{subfigure}[b]{0.325\linewidth}
		\includegraphics[width=\linewidth]{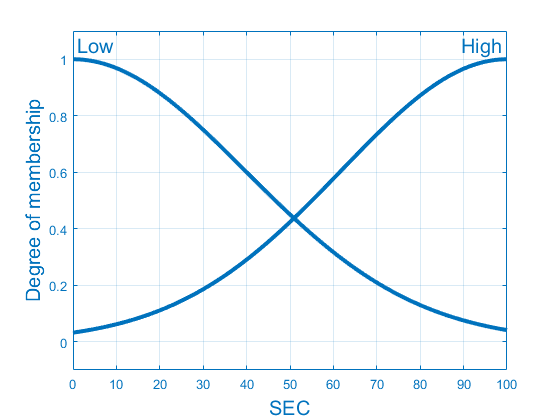}
		\caption{Membership functions for SEC input}
		\label{fig:fis1_sec}
	\end{subfigure}
	\begin{subfigure}[b]{0.325\linewidth}
		\includegraphics[width=\linewidth]{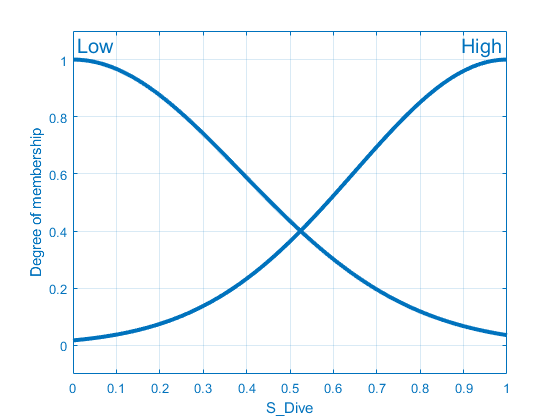}
		\caption{Membership functions for S-Dive output}
		\label{fig:fis1_sdive}
	\end{subfigure}
	\caption{Membership functions for inputs and outputs in FIS1 }
	\label{fig:fis1}
\end{figure*}

The first fuzzy inference system, known as FIS1, was designed to calculate S-Dive. This system is a Mamdani system and has two inputs called entropy and SEC, one output called S-Dive and two rules. All FIS1 membership functions are considered Gaussian functions. Figure \ref{fig:fis1} shows the membership functions for entropy, SEC, and S-Dive. It should be noted that the range for entropy is defined between 0 and 8, the range for SEC is defined between 0 and 100, and the range for S-Dive is defined between 0 and 1. The rules in this inference system are by

\begin{itemize}
	\item if Entropy is Low and SEC is High then S-Dive is Low
	\item if Entropy is High then S-Dive is High
\end{itemize}
\begin{figure*}[t]
	\centering
	\begin{subfigure}[b]{0.325\linewidth}
		\includegraphics[width=\linewidth]{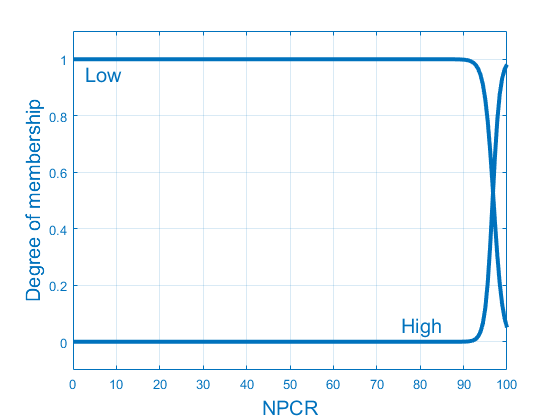}
		\caption{Membership functions for NPCR input}
		\label{fig:fis2_npcr}
	\end{subfigure}
	\begin{subfigure}[b]{0.325\linewidth}
		\includegraphics[width=\linewidth]{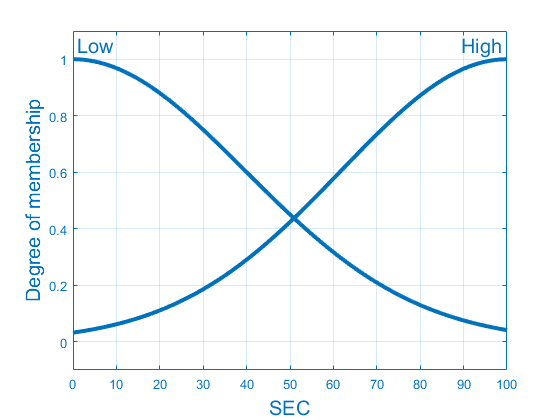}
		\caption{Membership functions for SEC input}
		\label{fig:fis2_sec}
	\end{subfigure}
	\begin{subfigure}[b]{0.325\linewidth}
		\includegraphics[width=\linewidth]{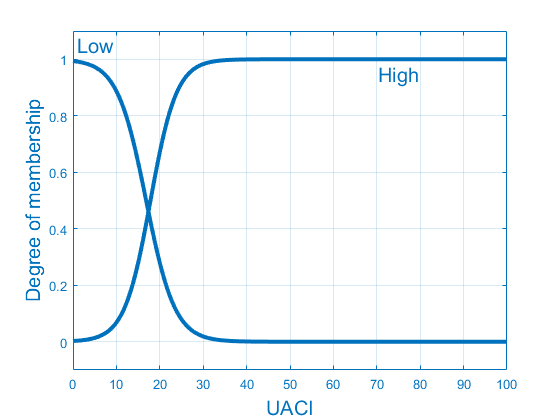}
		\caption{Membership functions for UACI input}
		\label{fig:fis2_uaci}
	\end{subfigure}
	\begin{subfigure}[b]{0.325\linewidth}
		\includegraphics[width=\linewidth]{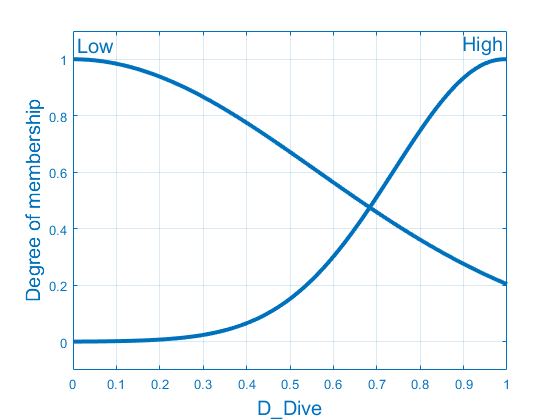}
		\caption{Membership functions for D-Dive output}
		\label{fig:fis2_ddive}
	\end{subfigure}
	\caption{Membership functions for inputs and outputs in FIS2}
	\label{fig:fis2}
\end{figure*}

FIS2 is also a Mamdani system. FIS2 has three inputs and one output and is designed to calculate D-Dive. The inputs of this system are the UACI, NPCR, and SEC, all of which are in the range between 0 and 100. The output of the system is D-Dive, which is in the range between 0 and 1. Figure \ref{fig:fis2} shows the membership functions for FIS2 inputs and outputs. In this system, all membership functions of Gaussian type are considered. FIS2 has three rules, which are stated below:
\begin{itemize}
	\item if UACI is Low and SEC is High then D-Dive is Low
	\item if NPCR is Low and SEC is High then D-Dive is Low
	\item if UACI is High and NPCR is High then D-Dive is High
\end{itemize}

In the AES-Chaos module, as in the process of calculating $x_{0}$, $x_{1}$, and $x_{2}$ from the primary key, the value of $x_{AES}$ is first calculated from the 2nd phase key by
\begin{equation}
	\label{eq:keyaes}
	x_{AES}=\dfrac{bi2de(2nd \; phase \; key)}{2^{128}},
\end{equation}
where '$bi2de$' converts a binary row vector to a non-negative decimal integer. The value $x_{AES}$  is used to create a pseudo-random sequence of natural numbers. Using the tent chaos map, a sequence is created and then sorted in ascending order. The obtained indices are used for encryption. After creating a random sequence of indices, the image is divided into $4 \times 4$ blocks. The first block is encrypted using the 2nd phase key by the AES algorithm. The encrypted content is used as the next block encryption key.

In the XOR by hash code module, the image is first divided into 512-bit blocks. The first block is XORed bit by bit with the code obtained in the first step of the first phase. The second block is also XORed bit by bit with the XOR result from the first block. In the same way, the next blocks are XORed bit by bit with the XOR result of their previous block.

\subsection{Decryption}
During the encryption process, a 1029-bit key called the decryption key is created. Each part of this key is intended as a system guide for the decryption process. This key consists of four parts, which consist of 512 bits as the primary key, 512 bits as the hash code, one bit as a flag to perform the AES-Chaos module process, and 4 bits as the number of XOR operations by hash code.

The primary key is used to calculate the values of $x_{0}$, $x_{1}$, $x_{2}$, $x_{AES}$, and the 2nd phase key based on equations \ref{eq:key} and \ref{eq:keyaes}. Using these values, XOR by tent chaotic map, pixel shuffling, genetic crossover, and AES-Chaos module can be performed in the reverse direction. Using the hash code, the XOR by hash code operation can be performed in the reverse direction. The last five bits of the decryption key are also used to guide the decryption system. If the bit that is the flag of the AES-Chaos module is one, it is necessary to perform the AES decryption operation on the encrypted image. Otherwise, there is no need to perform this operation. The value of the last four bits of the decryption key also indicates the number of XOR operations by hash code and is applied to the encrypted image to obtain the decrypted image. It is important to note that fuzzy inference systems are no longer used in the decryption process.

To secure the key, it is necessary to provide a secure channel for key transfer. Encrypted image security depends on the security of the encryption key. Due to the fact that the hashed value of the image is used in the decryption key, it is necessary to keep the decryption key confidential.

\section{Security Analysis}
\label{sec:result}
A suitable encryption algorithm should be robust against all existing cryptanalytic attacks
\cite{sheela2018image}.
In this section, we demonstrate the high security level of the proposed encryption algorithm by showcasing its performance according to a number of security analyses, including histogram, correlation analysis, key sensitivity test, security key space analysis, statistical analysis, information theory entropy, and differential analysis.

\subsection{Key Space and Brute Force Attack}
The key space of an encryption algorithm should be large enough to resist brute-force attacks
\cite{liu2019novel}.
Key length is 1029 bits in the proposed algorithm. The SHA-512 algorithm was used to extract a 512-bit key from the plain image for the XOR by hash code module. This method extended the key space significantly and increased key sensitivity to the master key with the SHA-512 algorithm. Therewith, as was said $x_{0}$, $x_{1}$, $x_{2}$, and $x_{2nd \; phase \; key}$ were obtained from the master key by SHA-512 and since the precision was $10^{-14}$, the key space was extended. Total key space was calculated with the following formula:
\begin{equation}
	S=2^{1029} \times 10^{14} \simeq 2^{1029} \times 2^{46}= 2^{1075}.
\end{equation}
which provides a safe key space.

\subsection{Histogram}
\begin{figure*}[t]
	\centering
	\begin{adjustbox}{width=\columnwidth,center}
	\begin{subfigure}[b]{0.25\linewidth}
		\includegraphics[width=\linewidth]{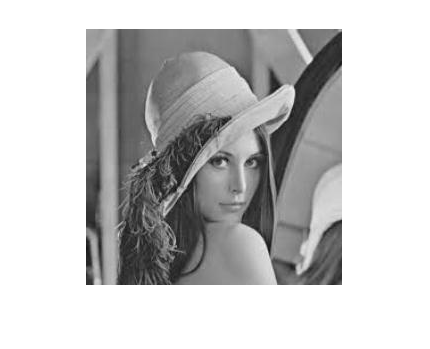}
		\caption{Lena Image}
		\label{fig:lena}
	\end{subfigure}
	\begin{subfigure}[b]{0.3\linewidth}
		\includegraphics[width=\linewidth]{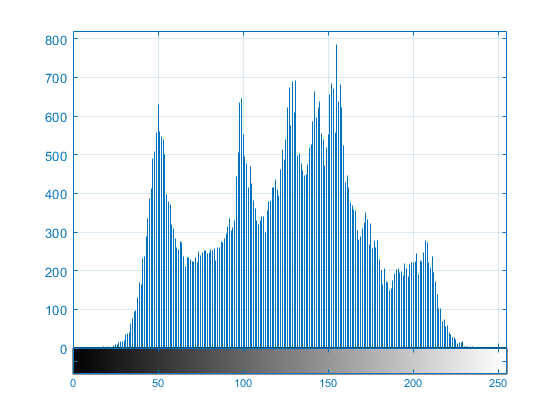}
		\caption{Histogram of Lena}
		\label{fig:lena_his}
	\end{subfigure}
	\begin{subfigure}[b]{0.25\linewidth}
		\includegraphics[width=\linewidth]{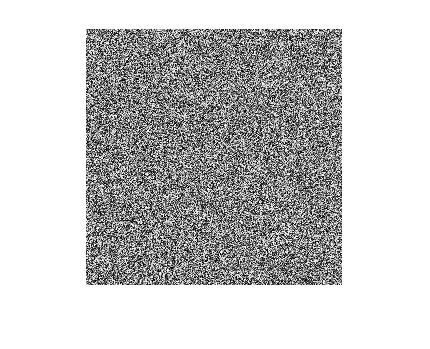}
		\caption{Lena encryption}
		\label{fig:lena_enc}
	\end{subfigure}
	\begin{subfigure}[b]{0.3\linewidth}
		\includegraphics[width=\linewidth]{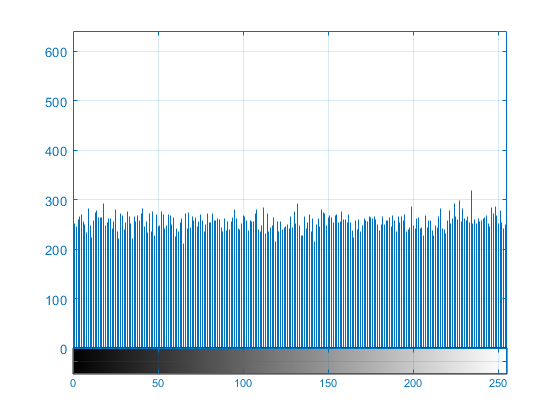}
		\caption{Histogram of Lena encryption}
		\label{fig:lena_enc_hist}
	\end{subfigure}
\end{adjustbox}
	\caption{Histogram of the image and histogram of the encrypted image}
	\label{fig:hist}
\end{figure*}
A histogram is a graph that shows the frequency of pixels with different values. In simpler term, a histogram is a bar chart whose horizontal axis represents the values of the pixels (light pixels on the right and dark pixels on the left) and its vertical axis represents the number of pixels in the image. It has already been stated that a good cryptographic system for images should be able to eliminate statistical features in images \cite{irani2019digital}. One of the best criteria for examining statistical features in plain and cipher images is to examine the histogram of these images. As can be seen in Figure \ref{fig:hist}, the histogram of the cipher image bears no resemblance to the plain image, and all statistical features of the input image are lost. An image encryption system should be able to create a smooth histogram for the encrypted image. As can be seen, the proposed system meets this requirement.

\subsection{Information Entropy}
Information entropy is a numerical measure of image randomness. The higher the entropy of an image, the greater the ambiguity about that image, and vice versa. Using the following formula, the entropy value of an image can be calculated \cite{mathivanan2019qr}:
\begin{equation}
	H(s) = \sum_{i=0}^{2^{M}-1} P(S_{i}) \log_2 \dfrac{1}{P(S_{i})}.
\end{equation}

In the above equation, $P(s_{i})$ represents the probability of $s_{i}$ and $2^{M}$ represents all states of information. The maximum value for the entropy of the encrypted image is 8. Table \ref{tab:ent} presents the entropy values for cipher images by different algorithms. By comparing these values and considering the calculated averages for encrypted images it can be concluded that the proposed system creates a better performance in creating ambiguity for encrypted images.

\begin{table}[t]
	\centering
	\caption{Entropy results}
	\begin{adjustbox}{width=\columnwidth,center}
	\begin{tabular}{lcccccccccc}
		& 	ours &
		Ref. \cite{wu2018image} &
		Ref. \cite{enayatifar2017image} &
		Ref. \cite{niyat2017color} &
		Ref. \cite{li2017image} &
		Ref. \cite{hermassi2013improvement} &
		Ref. \cite{ye2012efficient} &
		Ref. \cite{norouzi2014simple} &
		Ref. \cite{song2013image} &
		\makecell{AES in \\ CBC mode}\\ \hline
		lena      & 7.9974 & 7.9976 & 7.9975 & 7.9974 & 7.9913 & 7.9970 & 7.9970 & 7.9971 & 7.9972 & 7.9899\\
		cameraman & 7.9971 & 7.9975 & 7.9939 & 7.9971 & 7.9915 & 7.9972 & 7.9975 & 7.9974 & 7.9974 &7.9897 \\
		baboon    & 7.9972 & 7.9971 & 7.9938 & 7.9970 & 7.9912 & 7.9970 & 7.9972 & 7.9975 & 7.9971 &7.9900\\
		plain     & 7.9977 & 7.9970 & 7.9974 & 7.9970 & 7.9912 & 7.9973 & 7.9973 & 7.9966 & 7.9970 &7.9896\\
		ship      & 7.9972 & 7.9971 & 7.9941 & 7.9973 & 7.9907 & 7.9970 & 7.9971 & 7.9972 & 7.9971 &7.9889\\
		peppers   & 7.9976 & 7.9974 & 7.9958 & 7.9972 & 7.9909 & 7.9972 & 7.9974 & 7.9974 & 7.9971 &7.9900\\ \hline
		average   & 7.9974 & 7.9973 & 7.9954 & 7.9972 & 7.9911 & 7.9971 & 7.9972 & 7.9972 & 7.9972 &7.9897\\ \hline
	\end{tabular}
\end{adjustbox}
\label{tab:ent}
\end{table}

\subsection{Correlation of Pixels}

\begin{table}[t]
	\caption{Correlation coefficients of two adjacent pixels in plain and ciphered images}
	\begin{adjustbox}{width=\columnwidth,center}
	\begin{tabular}{llcccccccccc}
		&
		&
		ours &
		Ref. \cite{wu2018image} &
		Ref. \cite{enayatifar2017image} &
		Ref. \cite{niyat2017color} &
		Ref. \cite{li2017image} &
		Ref. \cite{hermassi2013improvement} &
		Ref. \cite{ye2012efficient} &
		Ref. \cite{norouzi2014simple} &
		Ref. \cite{song2013image} &
		\makecell{AES in \\ CBC mode}\\ \hline
		\multirow{3}{*}{lena}    & horizontal & 0.0062 & 0.0056 & 0.0023 & 0.0061 & 0.0041 & 0.0045 & 0.0199 & 0.0045 & 0.0294 &0.0143\\
		& vertical   & 0.0111 & 0.0037 & 0.0019 & 0.0116 & 0.0021 & 0.0344 & 0.0056 & 0.0370 & 0.0296 &0.0238\\
		& diagonal   & 0.0216 & 0.0032 & 0.0011 & 0.0018 & 0.0009 & 0.0097 & 0.0203 & 0.0098 & 0.0167 &0.0070\\ \hline
		\multirow{3}{*}{cameraman} &
		horizontal &
		0.0212 &
		0.0024 &
		0.0198 &
		0.0053 &
		0.0148 &
		0.0030 &
		0.0117 &
		0.0476 &
		0.0039 &0.0208\\
		& vertical   & 0.0217 & 0.0013 & 0.0132 & 0.0126 & 0.0084 & 0.0329 & 0.0137 & 0.0127 & 0.0033 &0.0124\\
		& diagonal   & 0.0186 & 0.0098 & 0.0032 & 0.0005 & 0.0026 & 0.0080 & 0.0395 & 0.0125 & 0.0215 &0.0088\\ \hline
		\multirow{3}{*}{baboon}  & horizontal & 0.0234 & 0.0026 & 0.0059 & 0.0060 & 0.0055 & 0.0011 & 0.0034 & 0.0470 & 0.0213 &0.0277\\
		& vertical   & 0.0043 & 0.0009 & 0.0041 & 0.0058 & 0.0015 & 0.0006 & 0.0209 & 0.0391 & 0.0090 &0.0077\\
		& diagonal   & 0.0081 & 0.0052 & 0.0028 & 0.0016 & 0.0041 & 0.0157 & 0.0151 & 0.0149 & 0.0053 &0.0089\\ \hline
		\multirow{3}{*}{plain}   & horizontal & 0.0029 & 0.0028 & 0.0062 & 0.0054 & 0.0075 & 0.0056 & 0.0148 & 0.0329 & 0.0347 &0.0143\\
		& vertical   & 0.0172 & 0.0041 & 0.0074 & 0.0081 & 0.0084 & 0.0039 & 0.0304 & 0.0639 & 0.0014 &0.0277\\
		& diagonal   & 0.0347 & 0.0010 & 0.0009 & 0.0021 & 0.0011 & 0.0090 & 0.0011 & 0.0615 & 0.0045 &0.0084\\ \hline
		\multirow{3}{*}{ship}    & horizontal & 0.0141 & 0.0001 & 0.0073 & 0.0085 & 0.0073 & 0.0041 & 0.0016 & 0.0234 & 0.0107 &0.0169\\
		& vertical   & 0.0215 & 0.0031 & 0.0109 & 0.0092 & 0.0216 & 0.0003 & 0.0099 & 0.0310 & 0.0062 &0.0166\\
		& diagonal   & 0.0196 & 0.0015 & 0.0016 & 0.0024 & 0.0035 & 0.0000 & 0.0153 & 0.0019 & 0.0121 &0.0320\\ \hline
		\multirow{3}{*}{peppers} & horizontal & 0.0040 & 0.0016 & 0.0037 & 0.0049 & 0.0021 & 0.0011 & 0.0160 & 0.0013 & 0.0154 &0.0219\\
		& vertical   & 0.0199 & 0.0059 & 0.0258 & 0.0031 & 0.0218 & 0.0006 & 0.0058 & 0.0084 & 0.0051 &0.0064\\
		& diagonal   & 0.0024 & 0.0034 & 0.0079 & 0.0079 & 0.0096 & 0.0157 & 0.0021 & 0.0095 & 0.0054 &0.0096\\ \hline
	\end{tabular}
\end{adjustbox}
\label{tab:corr}
\end{table}

\begin{figure*}[t]
	\captionsetup{justification=centering}
	\centering
	\begin{subfigure}[b]{0.23\linewidth}
		\includegraphics[width=\linewidth]{lena.png}
		\caption{}
		\label{fig:lena2}
	\end{subfigure}
	\begin{subfigure}[b]{0.23\linewidth}
		\includegraphics[width=\linewidth]{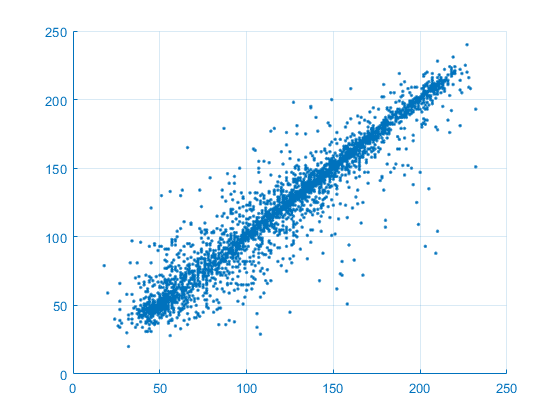}
		\caption{}
		\label{fig:corr_hor}
	\end{subfigure}
	\begin{subfigure}[b]{0.23\linewidth}
		\includegraphics[width=\linewidth]{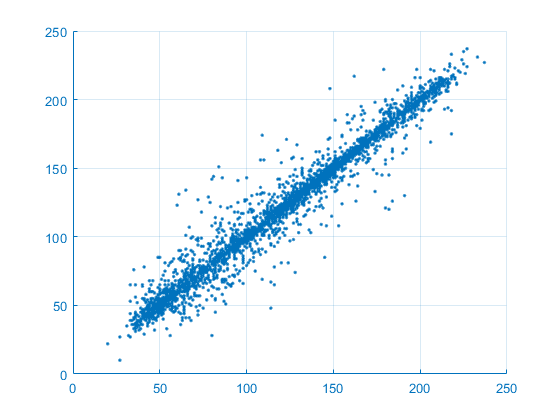}
		\caption{}
		\label{fig:corr_ver}
	\end{subfigure}
	\begin{subfigure}[b]{0.23\linewidth}
		\includegraphics[width=\linewidth]{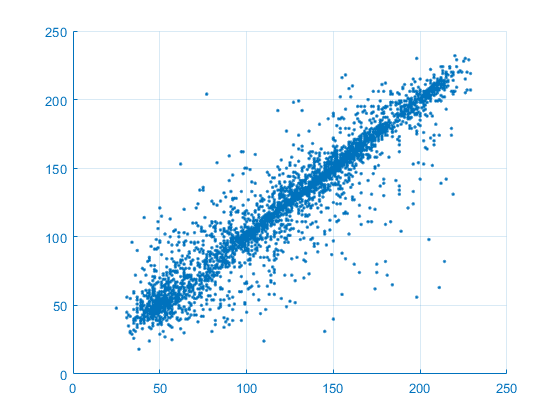}
		\caption{}
		\label{fig:corr_diag}
	\end{subfigure}
	\begin{subfigure}[b]{0.23\linewidth}
		\includegraphics[width=\linewidth]{lena_enc.png}
		\caption{}
		\label{fig:lena_enc2}
	\end{subfigure}
	\begin{subfigure}[b]{0.23\linewidth}
		\includegraphics[width=\linewidth]{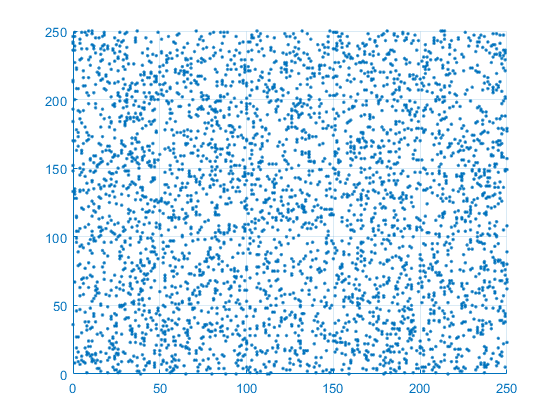}
		\caption{}
		\label{fig:corr_enc_hor}
	\end{subfigure}
	\begin{subfigure}[b]{0.23\linewidth}
		\includegraphics[width=\linewidth]{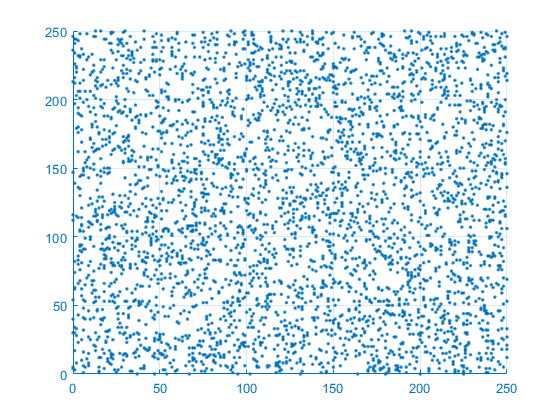}
		\caption{}
		\label{fig:corr_enc_ver}
	\end{subfigure}
	\begin{subfigure}[b]{0.23\linewidth}
		\includegraphics[width=\linewidth]{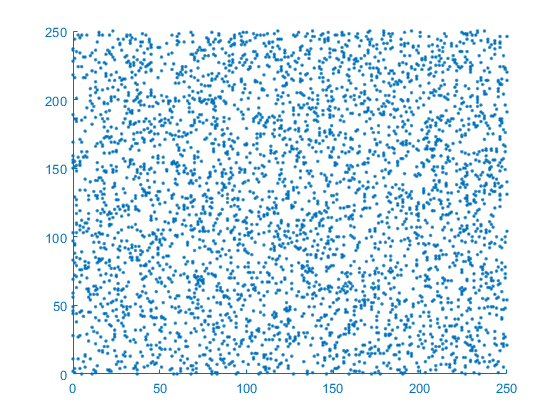}
		\caption{}
		\label{fig:corr_enc_diag}
	\end{subfigure}
	\caption{Correlation of adjacent pixels (a) Lena Image (b) Distribution of two horizontally adjacent pixels in the Lena image (c) Distribution of two vertically adjacent pixels in the Lena image (d)Distribution of two diagonally adjacent pixels in the Lena image (e) Lena encryption (f) Distribution of two horizontally adjacent pixels in the encrypted Lena image (g) Distribution of two vertically adjacent pixels in the encrypted Lena image (h) Distribution of two diagonally adjacent pixels in the encrypted Lena image}
	\label{fig:corr}
\end{figure*}

One of the most important targets in image encryption is to greatly decrease the correlation between adjacent pixels so that no relation between pixels could be spotted by naked eye. Calculating the correlation coeﬀicient, the formulas of which are shown below, is the most common method to measure this feature in horizontal, vertical, and diagonal directions \cite{ding2021deepkeygen}:

\begin{equation}
	\label{eq:corr}
	\begin{array}{l@{}l}
		r_{xy} = \dfrac{cov(x,y)}{\sqrt{D(x)D(y)}}, \\
		\\
		E_{x} = \dfrac{1}{N} \sum_{i=1}^{N} x_{i}, \\
		\\
		D_{x} = \dfrac{1}{N} \sum_{i=1}^{N} (x_{i}-E(x_{i}))^{2}, \\
		\\
		cov(x,y) = \dfrac{1}{N} \sum_{i=1}^{N} (x_{i}-E(x_{i}))(y_{i}-E(y_{i})).
	\end{array}
\end{equation}

The results of this factor have been represented in Table \ref{tab:corr}. Figure \ref{fig:corr} shows the correlation pixel diagram for the Lena image in horizontal, vertical, and diagonal modes. Figure \ref{fig:corr} also shows these diagrams for Lena's encrypted image. By comparing these diagrams, it can be concluded that the proposed system reliably eliminates the correlation of adjacent pixels in different directions.

\subsection{Differential Attacks}

\begin{table}[]
	\centering
	\caption{NPCR results}
	\begin{adjustbox}{width=\columnwidth,center}
		\begin{tabular}{lcccccccccc}
			& ours &
			Ref. \cite{wu2018image} &
			Ref. \cite{enayatifar2017image} &
			Ref. \cite{niyat2017color} &
			Ref. \cite{li2017image} &
			Ref. \cite{hermassi2013improvement} &
			Ref. \cite{ye2012efficient} &
			Ref. \cite{norouzi2014simple} &
			Ref. \cite{song2013image} &
			\makecell{AES in \\ CBC mode}\\ \hline
			lena      & 99.85 & 99.62 & 99.52 & 99.62 & 5.09E-04 & 35.76 & 99.21 & 99.73 & 99.64 &1.52E-03\\
			cameraman & 99.59 & 99.61 & 99.00 & 99.63 & 5.09E-04 & 97.15 & 99.25 & 99.70 & 99.59 &1.52E-03\\
			baboon    & 99.62 & 99.59 & 99.11 & 99.61 & 5.09E-04 & 89.60 & 99.15 & 99.67 & 99.60 &1.52E-03\\
			plain     & 99.58 & 99.62 & 99.42 & 99.65 & 5.09E-04 & 21.45 & 99.65 & 99.72 & 99.59 &1.52E-03\\
			ship      & 99.84 & 99.62 & 99.25 & 99.62 & 5.09E-04 & 78.49 & 99.26 & 99.72 & 99.62 &1.52E-03\\
			peppers   & 99.79 & 99.61 & 98.50 & 99.66 & 5.09E-04 & 57.89 & 99.17 & 99.68 & 18.96 &1.52E-03\\ \hline
			average   & 99.71 & 99.61 & 99.13 & 99.63 & 5.09E-04 & 63.39 & 99.28 & 99.70 & 86.16 &1.52E-03\\ \hline
		\end{tabular}
	\end{adjustbox}
	\label{tab:npcr}
\end{table}

\begin{table}[]
	\centering
	\caption{UACI results}
	\begin{adjustbox}{width=\columnwidth,center}
		\begin{tabular}{lcccccccccc}
			& ours &
			Ref. \cite{wu2018image} &
			Ref. \cite{enayatifar2017image} &
			Ref. \cite{niyat2017color} &
			Ref. \cite{li2017image} &
			Ref. \cite{hermassi2013improvement} &
			Ref. \cite{ye2012efficient} &
			Ref. \cite{norouzi2014simple} &
			Ref. \cite{song2013image} &
			\makecell{AES in \\ CBC mode}\\ \hline
			lena      & 33.52 & 33.42 & 33.59 & 33.42 & 1.9946E-06 & 0.28 & 33.43 & 33.33 & 33.52 &1.85E-04\\
			cameraman & 33.45 & 33.53 & 33.10 & 33.47 & 1.9946E-06 & 0.38 & 33.42 & 33.38 & 33.49 &7.77E-05\\
			baboon    & 33.63 & 33.38 & 33.25 & 33.41 & 1.9946E-06 & 0.35 & 33.38 & 33.15 & 33.50 &2.99E-05\\
			plain     & 33.54 & 33.64 & 33.53 & 33.51 & 1.9946E-06 & 0.17 & 33.62 & 33.62 & 33.50 &4.18E-05\\
			ship      & 33.59 & 33.66 & 33.59 & 33.51 & 1.9946E-06 & 1.24 & 33.46 & 33.30 & 33.32 &5.98E-06\\
			peppers   & 33.42 & 33.50 & 32.95 & 33.44 & 1.9946E-06 & 0.55 & 33.37 & 33.39 & 33.44 &1.79E-05\\ \hline
			average   & 33.53 & 33.52 & 33.33 & 33.46 & 1.9946E-06 & 0.50 & 33.45 & 33.36 & 33.46 &5.97E-05\\ \hline
		\end{tabular}
	\end{adjustbox}
	\label{tab:uaci}
\end{table}

In cryptography, differential attacks are common attacks where the plain image is changed in  as little as one bit, and then both the plain image and the changed image are encrypted and compared
\cite{shahna2020novel}.
Encryptor output must be sensitive to the change in plain image to resist such attacks. In other words, a slight change in plain image pixels affects the entire cipher image. Two common measurements for this property are the NPCR and UACI, which imply the number of pixel change  rate and the unified averaged changed intensity, respectively. In order to calculate these two quantities, plain image of $C_{1}$ changes to a pixel named $C_{2}$; $C_{1}$ and $C_{2}$ are then encrypted.

\subsubsection{NPCR}

The NPCR value is calculated using Equation \ref{eq:npcr} as follows \cite{torres2020exploring}:
\begin{equation}
	\label{eq:npcr}
	\begin{array}{l@{}l}
		NPCR(C_{1},C_{2}) = \dfrac{1}{M \times N} \sum_{i=1}^{M} \sum_{j=1}^{N} D(i, j), \\
		\\
		D(i ,j) = 
		\begin{cases}
			0  & C_{1}(i, j)=C_{2}(i, j)\\
			1  & C_{1}(i, j) \ne C_{2}(i, j)
		\end{cases}.
	\end{array}
\end{equation}
where $C_{1}(i,j)$ represents the pixel intensity of the encrypted grayscale image $C_{1}$, which is located in the $i th$ row and the $j th$ column. The results of the NPCR calculation for the proposed algorithm and several other algorithms are presented in Table \ref{tab:npcr}. Comparing the obtained results, it is observed that the proposed system has shown better performance than other algorithms.

\subsubsection{UACI}

As mentioned, another measure of the cryptographic system's resistance to differential attacks is the UACI. The UACI value is calculated using the following equation \cite{talhaoui2021fast}:
\begin{equation}
	UACI(C_{1},C_{2})=\dfrac{1}{M \times N} \sum_{i=1}^{M} \sum_{j=1}^{N} \dfrac{| C_{1}(i,j)-C_{2}(i,j) |}{255},
\end{equation}
where $C_{1}$ and $C_{2}$ are encrypted images. Table \ref{tab:uaci} presents the results of the UACI calculation for the proposed algorithm and 4 other algorithms. By close examination of the obtained values, the better performance of the proposed system can be verified.

\section{Conclusion}
\label{sec:conc}
In this paper, a new image encryption algorithm was proposed based on the AES-128 standard algorithm.
The proposed algorithm used a 1029-bit key. 
To increase key sensitivity, the master key was sent to the SHA-512 hash function, and then the 128-bit key was obtained for AES-128.
Also, the required initial conditions for the tent map were dependent upon this 128-bit key, which further increased the sensitivity.
Next, a random block was generated with a tent map the size of AES-128 encryption blocks ($4 \times 4$) which was XORed with the first plain image block before running the AES.
The sum of the last encrypted block affected the necessary initial value for the tent map to increase the diffusion property.
In this way, the permutation stage affected image pixels with only slight changes in the plain image, increasing resistance against differential attacks.





\end{document}